\documentclass[10pt,twocolumn,letterpaper]{article}

\usepackage{cvpr}              % To produce the CAMERA-READY version

\usepackage{booktabs}
\usepackage{siunitx}

\usepackage{threeparttable} % for \tnote and tablenotes
\definecolor{cvprblue}{rgb}{0.21,0.49,0.74}
\usepackage[pagebackref,breaklinks,colorlinks,allcolors=cvprblue]{hyperref}

\def\paperID{*****} % *** Enter the Paper ID here
\def\confName{}
\def\confYear{}

\title{Fluence Map Prediction with Deep Learning: A Transformer-based Approach}

\author{
Ujunwa Mgboh\textsuperscript{1} \quad 
Rafi Sultan\textsuperscript{1} \quad 
Dongxiao Zhu\textsuperscript{1} \quad 
Joshua Kim\textsuperscript{2}\\[4pt]
\textsuperscript{1}Wayne State University, Detroit, MI \quad 
\textsuperscript{2}Henry Ford Health, Detroit, MI\\[4pt]
{\tt\small \{ujunwa.mgboh,rafi.sultan,dzhu\}@wayne.edu} \quad 
{\tt\small jkim@hfhs.org}
}

\begin{document}
\maketitle
\begin{abstract}
\noindent\textbf{Background:} Accurate fluence map prediction is a critical step in intensity-modulated radiation therapy (IMRT), where the linear accelerator output is modulated to maximize tumor coverage while minimizing dose to adjacent healthy tissues. Fluence map optimization remains one of the most time-consuming and expertise-dependent components of radiotherapy treatment planning. To address this, we propose a novel deep learning framework that accelerates fluence map generation while ensuring consistent and high-quality treatment plans. \\[4pt]
\noindent\textbf{Methods:} We developed an end-to-end 3D Swin-UNETR that directly predicts nine-beam fluence maps from volumetric CT images and anatomical contours. This transformer-based network leverages hierarchical self-attention to capture both local anatomical structures and long-range spatial dependencies. The dataset comprised 99 prostate IMRT cases (79 for training, 20 for testing). Predicted fluence maps were imported into the Eclipse Treatment Planning System (Varian Medical Systems, Palo Alto, CA) for multileaf collimator sequencing and dose recalculation. Model performance was assessed using beam-wise fluence correlation, spatial gamma analysis, and DVH metrics. \\[4pt]
\noindent\textbf{Results:} The produced fluence maps yielding dose distributions in strong agreement with clinical reference plans. On the 20-patient test set, it achieved an average $R^2 = 0.95 \pm 0.02$, $\mathrm{MAE} = 0.035 \pm 0.008$, and gamma passing rates of $85 \pm 10\%$ (3\%/3~mm). No statistically significant differences were observed in DVH parameters for target volumes or organs at risk. The proposed Swin-UNETR framework enables fully automated fluence map prediction directly from anatomical inputs, eliminating reliance on precomputed dose information. By exploiting hierarchical self-attention for volumetric context modeling, the method enhances accuracy, spatial coherence, and efficiency in IMRT plan generation. These findings highlight its potential as a clinically scalable, inverse-free solution for automated radiotherapy planning.
\end{abstract}
    
\section{Introduction}
\label{sec:intro}

Radiotherapy uses high-energy radiation to target and kill cancer cells. The goal in all of radiation therapy treatment planning is to optimize the delivered dose to administer a tumorcidal dose of radiation to the targeted lesion while limiting the amount of dose given to surrounding healthy tissues. Within the field of radiation oncology, the most time-consuming step in the radiotherapy planning workflow has historically been the optimization of the treatment plan, which is an iterative process requiring several trials before the planner can achieve an optimal plan that best matches the radiation oncologist’s plan goals. Recently, the implementation of deep learning models in the treatment planning workflow has emerged as a potentially effective method for dramatically reducing the time required for plan generation and standardizing fluence map prediction accuracy and effectiveness, leveraging techniques used in fields like natural language processing and computer vision \cite{r1,r2,r3,r4}

Various input features are utilized in deep learning models for treatment plan automation tasks, influenced by factors like the specific radiotherapy application, input availability, and model selection. For instance, \cite{r7} enhanced gamma passing rate prediction and classification by combining dosiomics and plan features, leading to improved outcomes. Similarly, in predicting SBRT outcomes, a combination of dose distribution images and patient data enabled the identification of post-SBRT occurrence likelihood. In the context of fluence map prediction, input types such as image-based, dosimetric features, or machine learning-generated data are employed, aiming to optimize efficiency and accuracy in automated radiotherapy treatment planning \cite{r9,r10}.

Fluence map prediction, traditionally a labor-intensive process, has seen advancements through automated treatment planning (ATP) using deep learning models. Notably, a double convolutional neural network (CNN) incorporating patient anatomy and beam templates was employed to predict fluence maps for pancreas stereotactic body radiation therapy  \cite{r5}. Similarly, sequential CNN architectures tailored to unique input features and model designs for fluence prediction has been introduced \cite{r12,r13,r14}. Related studies primarily employed ResNet or UNet-based architectures. ResNet variants were used to enhance prediction accuracy \cite{r12} and to integrate knowledge-based planning for improved plan features \cite{r6}. UNet-based models, including both 2D and 3D variants, were applied for dose and fluence map prediction \cite{r14}, as well as for mapping dose distributions to fluence maps with demonstrated generalizability across cancer sites and radiotherapy techniques \cite{r17, r16}. These experiments further revealed insights into model performance concerning plan quality and delivery modalities, highlighting the potential of deep learning in fluence map prediction for radiotherapy planning.

The ability of deep learning to directly predict fluence/intensity maps without going through the time-consuming and disparity-prone inverse mapping process has been explored using multiple deep learning models \cite{r5,r6}. Building upon previous frameworks, this study extends fluence map prediction toward a comprehensive and clinically deliverable IMRT planning workflow for prostate cancer. We created a deep learning model from start to finish using the Swin UNETR architecture. This model uses hierarchical self-attention to pick up on both fine-grained anatomical details and long-range spatial dependencies in volumetric data. The model takes CT images and anatomical contours as inputs and uses them to predict nine volumetric fluence maps that match the beam intensities that can be delivered. The proposed framework learns the implicit relationship between anatomy and fluence within the network itself, eliminating the need for any external dose input. The predicted fluences are then imported into the treatment planning system for final dose calculation and DVH evaluation, the proposed method demonstrates the effectiveness of transformer-based architectures for accurate and clinically interpretable fluence prediction in radiotherapy planning.

\section{Materials and Methods}
\label{sec:formatting}

\subsection{Dataset Details}
 
In this research work, we utilized datasets from ninety-nine (99) prostate patients who underwent radiation therapy at our institution. Computed Tomography (CT) images were acquired using a Brilliance Big Bore (Philips Health Care, Cleveland, OH) scanner with the following parameters: 140 kVp, 500 mAs, 512x512 in-plane image dimensions, 1.28x1.28 ${mm}^2$ in-plane spatial resolution, and 3 mm slice thickness. Patients were originally treated using IMRT plans with 9 equally spaced fields to a total dose of 70.2-79.2 Gy in 1.8-2.0 Gy/fx. All plans were generated in the Eclipse TPS. Standard dose-volume histogram (DVH) metrics were utilized for target and organs-at-risk (OAR). Bladder, prostate, left and right femoral heads, and penile bulb were evaluated using criteria defined in QUANTEC guidelines \cite{r26}. Differences between the dose resulting from the predicted fluences and the original dose were evaluated using a paired two-sample t-test with a significance level of $p < 0.05$.
\subsection{Dataset Preprocessing}

The CT images, anatomical contours, and dose distributions for all patients were exported from the Eclipse Treatment Planning System (TPS) as DICOM files. The CT and contour data were processed using \textsc{MATLAB} (MathWorks, Natick, MA) scripts to generate structured arrays for each patient, while the fluence maps originally stored as text-based ``.optimal fluence'' files were extracted and converted into \texttt{NumPy} arrays for compatibility with the deep learning framework. All images were then resized to $128 \times 128$ pixels to maintain uniform spatial resolution across patients and modalities.

\subsection*{Input and Output Representation}

Each patient case was represented as a multi-channel volumetric tensor that integrates both anatomical and structural information. The CT and contour volumes were concatenated to form the model’s input tensor, allowing the network to jointly capture spatial relationships between anatomy and treatment structures. The model was trained in an end-to-end fashion to predict nine volumetric fluence maps, each corresponding to a distinct beam angle in a 9-field IMRT arrangement.

 The proposed model learns implicit dose–fluence relationships internally through its hierarchical attention mechanisms. This enables the network to infer dosimetric context directly from anatomical features without external dose supervision, ensuring a clinically practical and fully automated workflow.

\subsection*{Normalization}

All data modalities were independently min--max normalized to the range $[0, 1]$ using global minimum and maximum values determined from the training dataset. This normalization mitigated the effects of differing intensity scales among CT, contour and fluence data, thereby improving numerical stability and convergence during training.

\begin{table}
  \caption{\textbf{Twenty-case test set performance}}

  \label{tab:metric}
  \centering
  \begin{tabular}{@{}lc@{}}
    \toprule
    Metrics & {Mean $\pm$ SD} \\
    \midrule
    {$R^2$}                               & \num{0.95 \pm 0.02} \\
    MAE\tnote{a}                          & \num{0.035 \pm 0.008} \\
    RMSE\tnote{a}                         & \num{0.050 \pm 0.010} \\
    Mean $\Delta$Dose (\%)                & \num{-2.0 \pm 4.5} \\
    Gamma 3\%/3~mm pass rate (\%)         & \num{85 \pm 10} \\
    \bottomrule
  \end{tabular}
\end{table}

\section{Model Overview (End-to-End Framework)}
\label{sec:modeloverview}

\begin{figure*}[t]
  \centering
  \includegraphics[width=\textwidth]{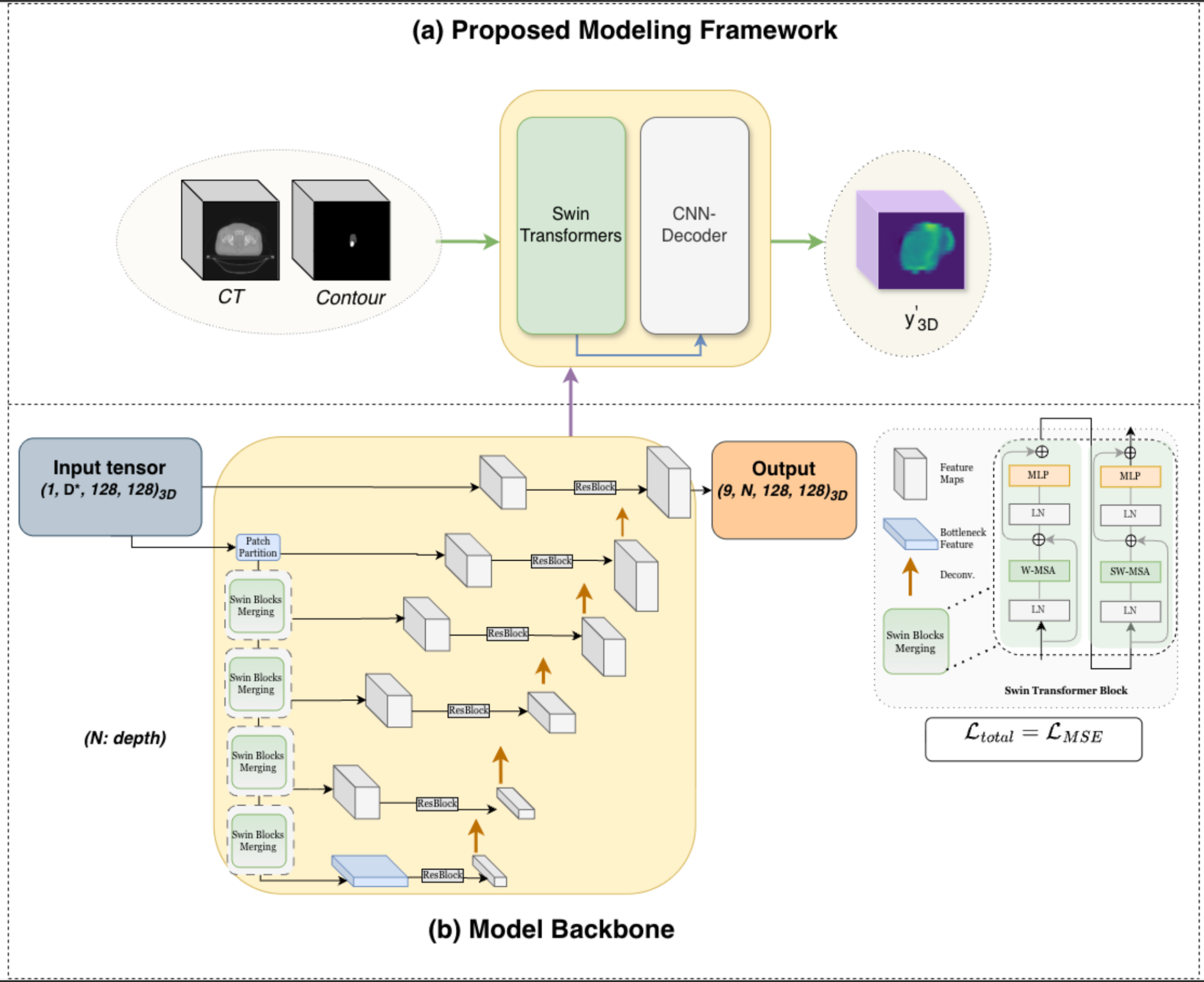}
  \caption{\textbf{Model architecture.} Overview of the proposed end-to-end deep learning framework for radiotherapy fluence map prediction. The model integrates volumetric CT images and contours into a unified input tensor and employs a 3D Swin-UNETR architecture to directly predict nine deliverable fluence maps corresponding to the clinical beam angles. The Swin-UNETR backbone combines hierarchical self-attention and UNet-style skip connections to capture both fine-grained anatomical structure and long-range spatial dependencies, enabling spatially coherent and clinically interpretable fluence predictions.}
  \label{fig:wide}
\end{figure*}

The proposed framework adopts an end-to-end deep learning design based on the Swin-UNETR architecture to predict deliverable fluence maps directly from patient anatomy (Figure~\ref{fig:wide}). 
The model integrates CT images and contours into a unified volumetric tensor and learns implicit dose–fluence relationships internally during training. 
This design preserves both local anatomical detail and global contextual dependencies through the Swin Transformer’s hierarchical self-attention mechanism, enabling accurate and spatially coherent fluence prediction without dependence on external dose priors.

The model was implemented in PyTorch using the MONAI framework. 
Training used the AdamW optimizer (\(\text{lr}=1\times10^{-4}\)) to improve computational efficiency. 
A batch size of 4 and 50 training epochs were used, with an 70/10/20 patient-level split for training, validation, and testing. 
Early stopping based on validation loss was applied to prevent overfitting and enhance generalization to unseen cases.
%-------------------------------------------------------------------------
\subsection{Swin-UNETR Architecture}
\label{sec:endtoend}

The input tensor for each patient consisted of two channels, the normalized CT volume and its corresponding contour mask of size $(2 \times N \times 128 \times 128)$, where \(N\) denotes the number of axial slices. 
The model outputs a nine-channel volumetric tensor $(9 \times N \times 128 \times 128)$, with each channel representing the predicted fluence map corresponding to one of the nine beam directions used in clinical 9-field IMRT planning. 
By operating on full 3D volumes, the model preserves anatomical continuity across slices and captures long-range spatial correlations essential for beam-level fluence modulation. 

The proposed framework employs a 3D Swin-UNETR architecture, which combines a hierarchical Swin Transformer encoder with a UNet-style decoder connected through residual skip connections. This design enables the network to integrate fine-grained local anatomy with global contextual awareness across the full planning volume.

The Swin Transformer backbone introduces hierarchical self-attention through shifted windowing, allowing the model to efficiently capture both local and long-range dependencies. This is particularly advantageous for fluence prediction, where the spatial distribution of intensity across beams depends on anatomical relationships that extend well beyond individual slices. The shifted-window mechanism ensures computational scalability with near-linear complexity, making it feasible to train on high-resolution 3D radiotherapy data without compromising spatial precision.

Meanwhile, the UNETR-style decoder facilitates multi-scale feature fusion, ensuring that anatomical detail and geometric alignment between CT anatomy, contours, and predicted fluence maps are preserved throughout the decoding process. This architecture offers improved robustness to inter-patient anatomical variation by combining global volumetric reasoning with voxel-level spatial fidelity, addressing the limitations of conventional 3D CNNs that rely on restricted receptive fields.

In summary, the Swin-UNETR framework effectively balances global context modeling and local spatial accuracy, making it well suited for automated fluence prediction in inverse-free IMRT planning.

\subsection{Loss Definition}

The model was trained to minimize the mean squared error (MSE) between the predicted fluence maps $\hat{F}_{b}(x)$ and the corresponding clinical ground-truth fluences $F_{b}(x)$ across all $B = 9$ beam channels and voxels $\Omega$:
\begin{equation}
\mathcal{L}_{\mathrm{MSE}}^{\mathrm{3D}} = 
\frac{1}{B\,|\Omega|} 
\sum_{b=1}^{9} 
\sum_{x \in \Omega} 
\big(\hat{F}_{b}(x) - F_{b}(x)\big)^{2}.
\end{equation}

This formulation enforces voxel-wise spatial consistency and penalizes large fluence deviations across beam directions. 
The resulting predicted fluences were post-processed and imported into the Eclipse Treatment Planning System (Varian Medical Systems, Palo Alto, CA) for multileaf collimator (MLC) sequencing, dose recalculation, and DVH-based evaluation of clinical deliverability.  

\begin{figure*}[t]
  \centering
  \ \includegraphics[width=\textwidth, trim=3mm 6mm 8mm 6mm, clip]{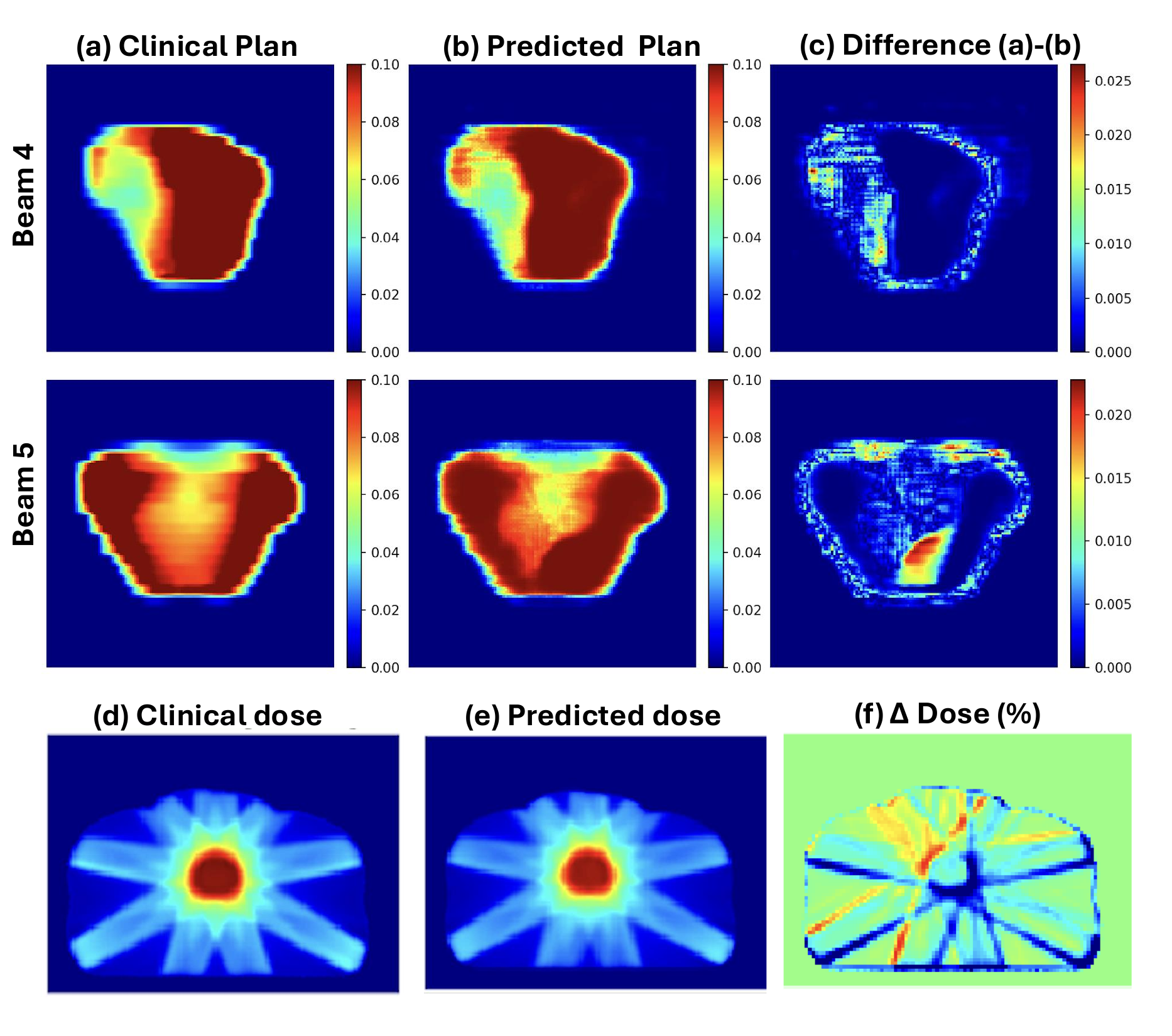}
  \caption{\textbf{Comparison of clinical and predicted dose distributions for representative test cases.}
  Panels (a–c) show beam-level fluence maps for two beams (Beam~4 and Beam~5), including 
  (a) the clinical plan, (b) the model-predicted plan, and (c) the voxel-wise difference 
  (clinical $-$ predicted). Panels (d–f) display axial composite dose distributions:
  (d) clinical, (e) predicted, and (f) percentage dose difference ($\Delta$Dose, \%). 
  The predicted doses exhibit high spatial and dosimetric agreement with the clinical reference.}
  \label{fig:combined_figure}
\end{figure*}

\section{Model Evaluation}
\label{sec:modeleval}

For model evaluation, 20 patients were excluded from training and validation. The CT volume of each patient and the corresponding contour set were combined into a two-channel volumetric tensor to form the input of the model, ensuring consistent anatomical and dosimetric alignment between the slices. The trained network generated nine fluence maps per patient, corresponding to the nine beam angles used in the clinical plans.

Following prediction, the fluence maps were post-processed, converted into deliverable beam parameters, and imported into the Eclipse Treatment Planning System (TPS) for dose recalculation using the same beam geometries, and machine parameters as the reference clinical plans. The recalculated dose distributions were exported from Eclipse as DICOM \texttt{RTDOSE} files for subsequent analysis.

Model performance was evaluated using both dosimetric and spatial agreement metrics. 
Dose–volume histograms (DVHs) were compared for all target and organ-at-risk (OAR) structures using indices such as $D_{95}$, $D_{98}$, and $V_{75}$. 
Voxel-wise agreement between predicted and clinical dose distributions was further quantified using statistical measures (coefficient of determination, mean absolute error, root-mean-square error, and mean percentage dose difference) and 3D gamma analysis under 3\%/3~mm criteria.

Together, these metrics provide a comprehensive assessment of the dosimetric trustworthiness and spatial consistency of the predicted dose distributions relative to the clinical reference plans.

\section{Results}
\label{sec:results}

\subsection{Quantitative Evaluation}
\label{sec:quantitative}

Quantitative analyses were performed across 20 test patients to evaluate both dosimetric and spatial agreement between the predicted and clinical dose distributions.

\begin{figure*}[t] 
\centering \includegraphics[width=\textwidth, trim=10mm 0mm 10mm 0mm, clip]{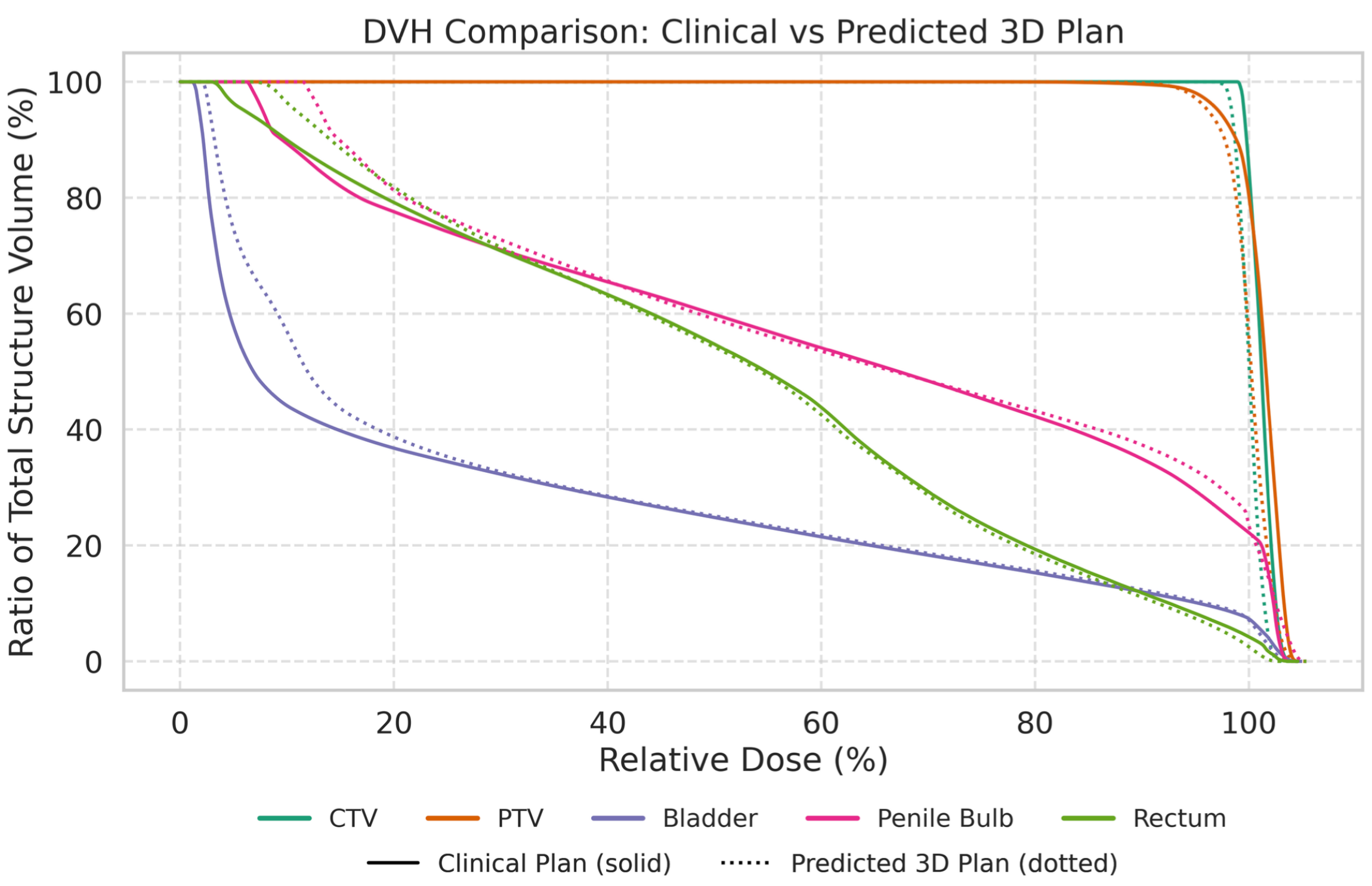} \caption{\textbf{Dose–volume histogram (DVH) for Patient~70.} The overlap shows significant similarity between the ground truth and predicted plan.} \label{fig_example2} 
\end{figure*}

\begin{table*}[t]
  \centering
  \small
  \setlength{\tabcolsep}{5pt} % adjust spacing between columns
  \caption{\textbf{Quantitative evaluation of the 20 test cases.}}
  \label{tab:quantitative}
  \vspace{2mm}
  \begin{tabular}{lcccc}
    \toprule
    \textbf{Structures} & \textbf{Dosage/Volume} & \textbf{Clinical Plan} & \textbf{Predicted Plan} & \textbf{P-Value} \\
    \midrule
    \textbf{CTV} & $D_{95}$ (Gy) & 87.27 $\pm$ 9.72 & 83.83 $\pm$ 8.23 & 0.39 \\
                 & $D_{98}$ (Gy) & 85.78 $\pm$ 8.53 & 82.12 $\pm$ 4.28 & 0.25 \\
    \midrule
    \textbf{PTV} & $D_{95}$ (Gy) & 89.27 $\pm$ 7.62 & 84.24 $\pm$ 2.53 & 0.51 \\
                 & $D_{98}$ (Gy) & 89.09 $\pm$ 7.64 & 82.53 $\pm$ 1.45 & 0.20 \\
    \midrule
    \textbf{LF}  & $D_{98}$ (Gy) & 36.95 $\pm$ 1.71 & 34.56 $\pm$ 2.70 & 0.44 \\
    \textbf{RF}  & $D_{98}$ (Gy) & 36.03 $\pm$ 2.24 & 35.61 $\pm$ 3.62 & 0.66 \\
    \midrule
    \textbf{Bladder} 
      & $V_{80}$ (\%) & 7.50 $\pm$ 5.76 & 9.04 $\pm$ 5.52 & 0.21 \\
      & $V_{75}$ (\%) & 15.23 $\pm$ 8.68 & 14.56 $\pm$ 6.51 & 0.44 \\
      & $V_{70}$ (\%) & 18.97 $\pm$ 10.20 & 17.95 $\pm$ 9.25 & 0.36 \\
      & $V_{65}$ (\%) & 22.41 $\pm$ 11.74 & 21.24 $\pm$ 8.89 & 0.64 \\
    \midrule
    \textbf{Rectum}
      & $V_{75}$ (\%) & 8.06 $\pm$ 2.87 & 8.47 $\pm$ 3.69 & 0.65 \\
      & $V_{70}$ (\%) & 13.17 $\pm$ 3.82 & 14.47 $\pm$ 3.55 & 0.34 \\
      & $V_{65}$ (\%) & 18.17 $\pm$ 4.61 & 19.78 $\pm$ 3.74 & 0.21 \\
      & $V_{60}$ (\%) & 23.24 $\pm$ 6.08 & 23.51 $\pm$ 3.27 & 0.26 \\
      & $V_{50}$ (\%) & 35.84 $\pm$ 6.08 & 38.67 $\pm$ 3.34 & 0.26 \\
    \midrule
    \textbf{Penile} & \textit{Min Dose (Gy)} & 11.19 $\pm$ 3.98 & 10.90 $\pm$ 2.14 & 0.46 \\
    \bottomrule
  \end{tabular}
\end{table*}

\subsubsection{DVH-Based Dosimetric Comparison}

Table~\ref{tab:quantitative} summarizes the statistical comparison of dose–volume indices for all evaluated structures. 
The predicted plans demonstrated strong dosimetric consistency with the clinical references. 
For both the clinical target volume (CTV) and planning target volume (PTV), differences in $D_{95}$ and $D_{98}$ were within 5\%, with $p > 0.2$ for all comparisons, indicating no statistically significant deviations. 
Organs-at-risk (OARs), including the bladder, rectum, and penile bulb, showed similar dose–volume parameters between predicted and clinical plans, confirming preservation of dose sparing to surrounding tissues. These results confirm that the proposed model achieves high quantitative agreement with the clinical plans across all key structures. The close alignment of DVH metrics demonstrates the model’s robustness and clinical reliability for consistent dose prediction in automated treatment planning.

\subsubsection{Voxel-Wise and Gamma Evaluation}
\label{sec:gammaeval}
To further assess spatial dose agreement, voxel-wise comparisons were performed between the predicted and clinical 3D dose distributions. 
In all cases, the mean performance was $R^2 = 0.95 \pm 0.02$, $\mathrm{MAE} = 0.035 \pm 0.008$, $\mathrm{RMSE} = 0.050 \pm 0.010$, and $\Delta\mathrm{Dose} = -2.0 \pm 4.5\%$, indicating a small global bias with excellent overall correlation. 
Three-dimensional gamma analysis yielded average passing rates of $85 \pm 10\%$ for the 3\%/3~mm criterion. 
Most discrepancies were localized in high-dose gradient regions, consistent with expected spatial uncertainties in dose recalculation.

Together, the DVH- and gamma-based evaluations confirm that the end-to-end Swin-UNETR framework produces dose distributions that are both clinically comparable and spatially consistent with the reference clinical plans.

\subsection{Qualitative Evaluation}
\label{sec:qualitative}

\subsubsection{DVH Visualization}
\label{sec:dvhvis}

Figure~\ref{fig_example2} presents representative dose–volume histograms comparing the clinical and predicted plans for a test patient. 
The predicted and clinical curves exhibit near-complete overlap across target volumes and OARs, reinforcing the quantitative results summarized in Table~\ref{tab:metric}.

\subsubsection{Dose Distribution and Fluence Map Comparison}
\label{sec:fluencevis}

Figure~\ref{fig:combined_figure} illustrates beam-wise qualitative comparisons between clinical and predicted fluence maps, alongside their corresponding difference maps. 
Additionally, axial dose slices (Figure~\ref{fig:combined_figure}(d)(e)(f)) display the clinical dose, predicted dose, and voxel-wise dose difference ($\Delta$Dose). 
The predicted fluence-derived dose distributions closely replicate the spatial intensity and modulation patterns of the clinical plans, with only minor localized deviations. 
These visual findings support the quantitative evidence that the model preserves both the dosimetric and spatial fidelity of the reference IMRT plans.

In summary, Table~\ref{tab:quantitative} indicates that the predicted plans generally align well with the ground truth, adhering to QUANTEC guidelines. This suggests that the predicted plans are reliable and effective in terms of dose distribution to both PTV volumes and organs at risk.

\section{Discussion and Conclusion}
\label{sec:discussion}

This study demonstrates the feasibility and efficiency of an end-to-end transformer-based deep learning framework for automatic fluence map prediction in prostate IMRT planning. 
The proposed 3D Swin-UNETR model directly learns the mapping from anatomical inputs, comprising CT images and contours, to deliverable fluence maps, achieving excellent dosimetric and spatial agreement with clinical reference plans. 
Across the test data, the model achieved a mean $R^2$ of $0.95 \pm 0.02$, $\mathrm{MAE} = 0.035 \pm 0.008$, and $\mathrm{RMSE} = 0.050 \pm 0.010$, with gamma passing rates of $85 \pm 10\%$, as seen in Table~\ref{tab:metric}. 
Predicted plans met clinical standards with no statistically significant differences in key DVH indices for both targets and organs-at-risk (Table~\ref{tab:quantitative}). 
In terms of computational efficiency, the model required only $3.97 \pm 0.25$~seconds per patient to generate all nine fluence maps, highlighting its potential to substantially accelerate the treatment planning workflow while maintaining clinical accuracy.

From a technical perspective, the transformer-based Swin-UNETR architecture offers several advantages over conventional convolutional neural networks (CNNs) that directly map patient anatomy to fluence or dose distributions~\cite{r5, r12, r14}. 
Whereas CNNs are inherently limited by local receptive fields and struggle to capture long-range dependencies, the hierarchical self-attention mechanism in Swin-UNETR enables simultaneous modeling of fine anatomical detail and global contextual relationships across the entire 3D volume. 
This improved representational capacity enhances the model’s ability to capture dose–fluence interactions and structural correlations between distant anatomical regions, such as between the bladder, rectum, and prostate in pelvic radiotherapy.

The end-to-end design also improves clinical practicality and reduces reliance on intermediate dose predictions, aligning the workflow more closely with real-world automated planning objectives. 
By learning the implicit relationships between anatomy, dose, and beam fluence directly within a single model, the framework simplifies implementation, minimizes error propagation, and ensures consistent output quality across patient cases. 
The results demonstrate that such transformer-based models can deliver accurate and spatially coherent fluence predictions suitable for direct integration into clinical treatment planning systems.

The drawback in this work is that it uses only prostate IMRT cases, which typically exhibit less anatomical and modulation variability than more complex treatment sites such as the head and neck. 
Secondly, although the two-stage architecture reduces error through hierarchical learning, the model’s generalizability remains constrained by the relatively small dataset size and the single-institution scope of testing.

Future work will extend this framework to multi-institutional and multi-modality datasets to enable broader validation across anatomically complex regions. 
Cross-institutional studies will be conducted to evaluate robustness under variations in contouring protocols, beam configurations, and dose calculation algorithms, thereby enhancing the clinical generalizability and scalability of the proposed method.

\section{Acknowledgments}
I am also thankful to Thomas Rumble Fellowship provided by Wayne State University and Henry Ford Health for providing the necessary financial support, without which this research would not have been possible.

{
    \small
\bibliographystyle{ieeenat_fullname}

}


\end{document}